\documentclass[12pt,letterpaper]{article}
\usepackage{amsmath,amsfonts,amsthm,amssymb}

\theoremstyle{plain}

\numberwithin{equation}{section}
\newtheorem{thm}{Theorem}[section]
\newtheorem{lem}[thm]{Lemma}

\newenvironment{exam}[1]{\medskip%
  \setlength{\rightmargin}{\leftmargin}%
  {\noindent\textbf{Example #1}.\enspace}%
  }%

\newcounter{cond}

\newcommand{\complex}{{\mathbb C}}

\newcommand{\real}{{\mathbb R}}
\newcommand{\integers}{{\mathbb Z}}
\newcommand{\ascript}{{\mathcal A}}

\newcommand{\cupdot}{\mathbin{\cup{\hskip-5.4pt}^\centerdot}\,}
\newcommand{\bigcupdotionen}{{\bigcup _{i=1}^n}{\hskip-8pt}^\centerdot{\hskip 8pt}}
\newcommand{\bigcupdotitwon}{{\bigcup _{i=2}^n}{\hskip-8pt}^\centerdot{\hskip 8pt}}
\newcommand{\bigcupdotijn}{{\bigcup _{i=j}^n}{\hskip-8pt}^\centerdot{\hskip 8pt}}

\newcommand{\bigcupdotijonen}{{\bigcup _{i=j+1}^n}{\hskip-13pt}^\centerdot{\hskip 13pt}}

\newcommand{\subcupdot}{\mathbin{\cup{\hskip-4pt}^\centerdot}\,}

\newcommand{\ab}[1]{\left|#1\right|}
\newcommand{\brac}[1]{\left\{#1\right\}}
\newcommand{\paren}[1]{\left(#1\right)}
\newcommand{\sqbrac}[1]{\left[#1\right]}

\begin{document}

\title{QUANTUM REALITY FILTERS}
\author{Stan Gudder\\ Department of Mathematics\\
University of Denver\\ Denver, Colorado 80208\\
sgudder@math.du.edu}
\date{}
\maketitle

\begin{abstract}
An anhomomorphic logic $\ascript ^*$ is the set of all possible realities for a quantum system. Our main goal is to find the ``actual reality'' $\phi _a\in\ascript ^*$ for the system. Reality filters are employed to eliminate unwanted potential realities until only $\phi _a$ remains. In this paper, we consider three reality filters that are constructed by means of quantum integrals. A quantum measure $\mu$ can generate or actualize a $\phi\in\ascript ^*$ if $\mu (A)$ is a quantum integral with respect to $\phi$ for a density function $f$ over events $A$. In this sense, $\mu$ is an ``average'' of the truth values of $\phi$ with weights given by $f$. We mainly discuss relations between these filters and their existence and uniqueness properties. For example, we show that a quadratic reality generated by a quantum measure is unique. In this case we obtain the unique actual quadratic reality. 

\medskip
\noindent ``Reality is merely an illusion, albeit a very persistent one.''

{\hfill ---{\footnotesize Albert Einstein}}
\end{abstract}

\section{Introduction}  
In the past, one of the main goals of physics has been to describe physical reality. Recently however, physicists have embarked on the even more ambitious program of actually \textit{finding} physical reality. Specifically, their quest is to find the universal truth function $\phi _a$. If $A$ is any proposition concerning the physical universe, then
$\phi _a(A)$ is $0$ or $1$ depending on whether $A$ is false or true. Assuming that the set of propositions is a Boolean algebra $\ascript$, we can think of $\ascript$ as an algebra of subsets of a universe $\Omega$ of outcomes. The outcomes are frequently interpreted as paths (or trajectories or histories) of a physical world. The actual physical universe $\Omega _1$ is vast, complicated and probably infinite. To make the situation more manageable, we shall only consider a toy universe $\Omega$ with a finite number of elements. This will still enable us to investigate structures that may be applicable to $\Omega _1$.

The theory that we present originated with the work of R.~Sorkin \cite{sor941, sor942} who was motivated by the histories approach to quantum mechanics and quantum gravity and cosmology \cite{hal09, mocs05, sor07}. Sorkin called this quantum measure theory and anhomomorphic logic. After Sorkin's pioneering work, many investigators have developed various aspects of the theory \cite{dgt08,gt09,gtw09,gud10,gudms,gud091,gud092,gud093, sal02, sor09, sw08}. For a classical universe, a truth function would be a homomorphism $\phi$ from the algebra $\ascript$ of propositions or events to the two-element Boolean algebra $\integers _2=\brac{0,1}$. However, because of quantum interference, a truth function describing a quantum reality need not be a homomorphism. We call the set
\begin{equation*}
\ascript ^*=\brac{\phi\colon\ascript\to\integers _2\colon\phi (\emptyset )=0}
\end{equation*}
the \textit{full anhomomorphic logic}. The elements of $\ascript ^*$ are interpreted as potential realities for a physical system and are called \textit{coevents}. Our task is to find the actual reality $\phi _a\in\ascript ^*$ which describes what actually happens.

Even if the cardinality $n=\ab{\Omega}$ of $\Omega$ is small, the cardinality $\ab{\ascript ^*}=2^{(2^n-1)}$ of
$\ascript ^*$ can be immense. Hence, it is important to establish reality filters that filter out unwanted potential realities until we are left with the actual reality $\phi _a$. Mathematically, reality filters are requirements that can be employed to distinguish $\phi _a$ from other possible coevents. One of the main reality filters that has been used is called preclusivity \cite{gt09, gtw09, sor941, sor942, sor09}. Nature has provided us with a quantum measure $\mu$ which is related to the state of the system. The measure $\mu$ is where the physics is contained and information about $\mu$ is obtained by observing the physical universe. Specifically, $\mu$ is a nonnegative set function on $\ascript$ that is more general than an ordinary measure. For $A\in\ascript ^*$, $\mu (A)$ is interpreted as the propensity that the event $A$ occurs. If $\mu (A)=0$, then $A$ does not occur and we say that $A$ is \textit{precluded}. We say that
$\phi\in\ascript ^*$ is \textit{preclusive} if $\phi (A)=0$ for all precluded $A\in\ascript$. It is generally agreed that
$\phi _a$ should be preclusive.

Although preclusivity is an important reality filter, it is too weak to determine $\phi _a$ uniquely. Other reality filters that have been used involve the algebraic properties of coevents and are called unital, additive, multiplicative and quadratic properties \cite{gt09, gud092, gud093, sor941, sor942, sor09}. Unfortunately, there does not seem to be agreement on which, if any, of these properties is appropriate. In this paper we shall consider three other reality filters, two of which were proposed in \cite{gud093}. These filters involve coevents that determine the quantum measure using an averaging process called a quantum integral. These filters are called 1-generated, 2-generated and actualized.

The present paper continues our study of 1-generated and 2-generated coevents and introduces the concept of an actualized coevent. One of our results shows that if $\phi$ and $\psi$ are 1-generated by the same quantum measure, then $\phi =\psi$. Another result shows that if $\phi$ and $\psi$ are quadratic coevents that are 2-generated by the same quantum measure, then $\phi =\psi$. It follows that the 1- and 2-generated filters uniquely determine a quadratic reality. We also demonstrate that this result does not hold for actualized coevents. Still another result shows that 1-generated coevents are 2-generated and that unital 1-generated coevents are actualized. We give examples of coevents that are 2-generated and actualized but are not 1-generated. We also give an example of a coevent that is actualized but we conjecture is not 2-generated.

Quantum measures that 1- or 2-generate a coevent appear to belong to a rather restricted class. The main reason for introducing the actualized filter is that it appears to be more general so it admits a larger set of actualizing quantum measures. We present evidence of this fact by considering a sample space $\Omega$ with $\ab{\Omega}=2$. Various open problems are presented. These problems mainly concern quantum measures and coevents that correspond to our three reality filters. We begin an approach to one of these problems by characterizing 1-generated coevents.

\section{Anhomomorphic Logic \\ and Quantum Integrals} 
Let $\Omega$ be a finite nonempty set with cardinality $\ab{\Omega}=n$. We call $\Omega$ a
\textit{sample space}. We think of the elements of $\Omega$ as outcomes or trajectories of an experiment or physical system and the collection of subsets $\ascript =2^\Omega$ of $\Omega$ as the possible events. We can also view the elements of $\ascript$ as propositions concerning the system. Contact with reality is given by a truth function $\phi\colon\ascript\to\integers _2$ where $\integers _2=\brac{0,1}$ is the two-element Boolean algebra with the usual multiplication and addition given by $0\oplus 0=1\oplus 1=0$ and $0\oplus 1=1\oplus 0=1$.

For $\omega\in\Omega$ we define \textit{evaluation map} $\omega ^*\colon\ascript\to\integers _2$ given by
\begin{equation*}
\omega ^*(A)=
\begin{cases}
 1&\text{if }\omega\in A\\
 0&\text{if }\omega\notin A
\end{cases}
\end{equation*}
For classical systems, it is assumed that a truth function $\phi$ is a homomorphism; that is, $\phi$ satisfies
\begin{list} {(H\arabic{cond})}{\usecounter{cond}
\setlength{\rightmargin}{\leftmargin}}
\item $\phi (\Omega )=1$\enspace (unital)
\item $\phi (A\cupdot B)=\phi (A)\oplus\phi (B)$ whenever $A\cap B=\emptyset$\enspace (additivity)
\item $\phi (A\cap B)=\phi (A)\phi (B)$\enspace (multiplicativity)
\end{list}
In (H2) $A\cupdot B$ denotes $A\cup B$ whenever $A\cap B=\emptyset$. It is well-known that $\phi$ is a homomorphism if and only if $\phi =\omega ^*$ for some $\omega\in\Omega$. Thus, there are $n$ truth functions for classical systems.

As discussed in \cite{gt09, sor941, sor942, sor09}, for a quantum system a truth function need not be a homomorphism. Various conditions for quantum truth functions have been proposed. In \cite{sor941, sor942} it is assumed that quantum truth functions satisfy (H2) and these are called \textit{additive} truth functions while in
\cite{gt09, sor09} it is assumed that quantum truth functions satisfy (H3) and these are called \textit{multiplicative} truth functions. In \cite{gud092} it is argued that quantum truth functions need not satisfy (H1), (H2) or (H3) but should be \textit{quadratic} or \textit{grade}-2 \textit{additive} \cite{gt09, sor942} in the sense that\medskip

\noindent (Q4)\enspace $\phi (A\cupdot B\cupdot C)=\phi (A\cupdot B)\oplus\phi (A\cupdot C)
  \oplus\phi (B\cupdot C)\oplus\phi (A)\oplus\phi (B)\oplus\phi (C)$
\medskip

If $\phi ,\psi\colon\ascript\to\integers _2$ we define $\phi\psi\colon\ascript\to\integers _2$ by
$(\phi\psi )(A)=\phi (A)\psi (A)$ and $\phi\oplus\psi\colon\ascript\to\integers _2$ by
$(\phi\oplus\psi )(A)=\phi (A)\oplus\psi (A)$ for all $A\in\ascript$. We define the $0$ and $1$ truth functions by $0(A)=0$ for all $A\in\ascript$ and $1(A)=1$ if and only if $A\ne\emptyset$. It can be shown
\cite{gt09, gud092, sor941} that $\phi$ is additive if and only if $\phi$ is a degree-1 polynomial
\begin{equation*}
\phi =\omega _1^*\oplus\cdots\oplus\omega _m^*
\end{equation*}
and that $\phi$ is multiplicative if and only if $\phi$ is a monomial
\begin{equation*}
\phi =\omega _1^*\omega _2^*\cdots\omega _m^*
\end{equation*}
Moreover, one can show \cite{gt09, gud092} that $\phi$ is quadratic if and only if $\phi$ is a degree-1 polynomial or
$\phi$ is a degree-2 polynomial of the form
\begin{equation*}
\phi =\omega _1^*\oplus\cdots\oplus\omega _m^*\oplus\omega _i^*\omega _j^*
  \oplus\cdots\oplus\omega _r^*\omega _s^*
\end{equation*}
We call $\ascript ^*=\brac{\phi\colon\ascript\to\integers _2\colon\phi (\emptyset )=0}$ the full
\textit{anhomomorphic logic} and the elements of $\ascript ^*$ are called \textit{coevents}. There are $2^{(2^n-1)}$ coevents of which $n$ are classical, $2^n-1$ are additive, $2^n-1$ are multiplicative and $2^{n(n+1)/2}$ are quadratic. It can be shown that any coevent can be written as a polynomial in the evaluation maps and that such an evaluation map representation is unique up to the order of its terms \cite{gt09, gud092}.

Applying (Q4) one can prove by induction that $\phi\in\ascript ^*$ is quadratic if and only if
\begin{equation}         
\label{eq21}
\phi\paren{A_1,\cupdot\cdots\cupdot A_m}=\bigoplus _{i<j=1}^m\phi (A_i\cupdot A_j)
  \oplus\tfrac{1}{2}\sqbrac{1-(-1)^m}\bigoplus _{i=1}^m\phi (A_i)
\end{equation}
for all $m\ge 3$. It follows from \eqref{eq21} that a quadratic coevent is determined by its values on singleton and doubleton sets in $\ascript$. Moreover, given any assignment of zeros and ones to the singleton and doubleton sets in $\ascript$, there exists a unique quadratic coevent that has these values.

Following \cite{gud091}, if $f\colon\Omega\to\real$ and $\phi\in\ascript ^*$, we define the $q$-\textit{integral}
\begin{equation*}
\int fd\phi =\int _0^\infty\phi\paren{\brac{\omega\colon f(\omega )>\lambda}}d\lambda
  -\int _0^\infty\phi\paren{\brac{\omega\colon f(\omega )<-\lambda}}d\lambda
\end{equation*}
where $d\lambda$ denotes Lebesgue measure on $\real$. If $f\ge 0$, then
\begin{equation}         
\label{eq22}
\int fd\phi =\int _0^\infty\paren{\brac{\omega\colon f(\omega )>\lambda}}d\lambda
\end{equation}
and we shall only integrate nonnegative functions here. Denoting the characteristic function of a set $A$ by $\chi _A$, any $f\colon\Omega\to\real ^+$ has the canonical representation $f=\sum _{i=1}^n\alpha _i\chi _{A_i}$ where
$0<\alpha _1<\cdots <\alpha _n$ and $A_i\cap A_j=\emptyset$, $i\ne j$. It follows from \eqref{eq22} that
\begin{align}         
\label{eq23}
\int fd\phi &=\alpha _1\phi\paren{\bigcupdotionen A_i} + (\alpha _2-\alpha _1)
  \phi\paren{\bigcupdotitwon A_i}+\cdots +(\alpha _n-\alpha _{n-1})\phi (A_n)\notag\\\noalign{\smallskip}
&=\sum _{j=1}^n\alpha _j\sqbrac{\phi\paren{\bigcupdotijn A_i}-\phi\paren{\bigcupdotijonen A_i}}
\end{align}
It is clear from \eqref{eq22} or \eqref{eq23} that $\int fd\phi\ge 0$ if $f\ge 0$. Also, it is easy to check that
$\int\alpha fd\phi =\alpha\int fd\phi$ for all $\alpha\in\real$. However, the $q$-integral is not linear because
\begin{equation*}
\int (f+g)d\phi\ne\int fd\phi +\int gd\phi
\end{equation*}
in general. Moreover, in general we have
\begin{equation*}
\int fd(\phi\oplus\psi )\ne\int fd\phi +\int fd\psi
\end{equation*}
As usual in integration theory, for $A\in\ascript$ and $\phi\in\ascript ^*$ we define
\begin{equation*}
\int _Afd\phi =\int f\chi _Ad\phi
\end{equation*}
In general,
\begin{equation*}
\int _{A\subcupdot B}fd\phi\ne\int _Afd\phi +\int _Bfd\phi
\end{equation*}
The $q$-integral is not even grade-2 additive because, in general
\begin{equation*}
\int _{A\subcupdot B\subcupdot C}\!\!fd\phi\ne\int _{A\subcupdot B}fd\phi +\int _{A\subcupdot C}fd\phi
  +\int _{B\subcupdot C}fd\phi -\int _Afd\phi -\int _Bfd\phi -\int _Cfd\phi
\end{equation*}

\section{Reality Filters} 
This section introduces the three quantum reality filters discussed in the introduction. A $q$-\textit{measure} is a set function $\mu\colon\ascript\to\real ^+$ that satisfies the \textit{grade}-2 \textit{additivity condition}
\begin{equation}         
\label{eq31}
\mu\paren{A\cupdot B\cupdot C}=\mu (A\cupdot B)+\mu (A\cupdot C)+\mu (B\cupdot C)-\mu (A)-\mu (B)-\mu (C)
\end{equation}
Condition \eqref{eq31} is more general than the usual (\textit{grade}-1) \textit{additivity}
$\mu (A\cupdot B)=\mu (A)+\mu (B)$ for measures. A $q$-measure $\mu$ is \textit{regular} if it satisfies
\begin{list} {(R\arabic{cond})}{\usecounter{cond}
\setlength{\rightmargin}{\leftmargin}}
\item $\mu (A)=0$ implies $\mu (A\cupdot B)=\mu (B)$.
\item $\mu (A\cupdot B)=0$ implies $\mu (A)=\mu (B)$.
\end{list}
It is frequently assumed that a $q$-measure is regular but, for generality, we shall not make that assumption here. An example of a regular $q$-measure is a map $\mu (A)=\ab{\nu (A)}^2$ where $\nu\colon\ascript\to\complex$ is a complex measure. Of course, complex measures arise in quantum mechanics as amplitude measures. A more general example of a regular $q$-measure is a decoherence functional that is employed in the histories approach to quantum mechanics \cite{hal09, mocs05, sor942}. A $q$-measure $\mu$ is determined by its values on singleton and doubleton sets because it follows from \eqref{eq31} and induction that
\begin{equation}         
\label{eq32}
\mu\paren{\brac{\omega _1,\ldots ,\omega _m}}=\sum _{i<j=1}^m\mu\paren{\brac{\omega _i,\omega _j}}
  -(m-2)\sum _{i=1}^m\mu (\omega _i)
\end{equation}
for all $m\ge 3$, where $\mu (\omega _i)$ is shorthand for $\mu\paren{\brac{\omega _i}}$.

We assume that nature provides us with a fixed $q$-measure $\mu\colon\ascript\to\real ^+$ and that $\mu (A)$ can be interpreted as the propensity that $A$ occurs. If $\mu (A)=0$, then $A$ does not occur and $A$ is
$\mu$-\textit{precluded}. We say that $\phi\in\ascript ^*$ is $\mu$-\textit{preclusive} if $\phi (A)=0$ whenever
$\mu (A)=0$. The $q$-measure $\mu$ 1-\textit{generates} $\phi\in\ascript ^*$ if there exists a strictly positive function $f\colon\Omega\to\real$ such that $\mu (A)=\int _Afd\phi$ for all $A\in\ascript$. We call $f$ a $\phi$-\textit{density} for
$\mu$. Thus, $\mu$ is an ``average'' over the truth values of $\phi$ weighted by the density $f$.

Unfortunately, there are many $q$-measures that do not 1-generate any coevent. One reason for this is that when $\ab{\Omega}=n$, then $f\colon\Omega\to\real$ gives at most $n$ pieces of information, while a $q$-measure is determined by its values on singleton and doubleton sets so $n(n+1)/2$ pieces of information may be needed. 
We therefore introduce a more complicated (and as we shall show, more general) definition. A function
$f\colon\Omega\times\Omega\to\real$ is \textit{symmetric} if $f(\omega ,\omega ')=f(\omega ',\omega )$ for all
$\omega ,\omega '\in\Omega$. Notice that a symmetric function on $\Omega\times\Omega$ has $n(n+1)/2$ possible values. A $q$-measure $\mu$ on $\ascript$ 2-\textit{generates} $\phi\in\ascript ^*$ if there exists a strictly positive symmetric function $f\colon\Omega\times\Omega\to\real$ such that
\begin{equation*}
\mu (A)=\int _A\sqbrac{\int _Af(\omega ,\omega ')d\phi (\omega )}d\phi (\omega ')
\end{equation*}
for every $A\in\ascript$. Again, $f$ is a $\phi$-\textit{density} for $\mu$. It can be shown that if $\phi$ is
1- or 2-generated by $\mu$, then $\phi$ is $\mu$-preclusive \cite{gud093}.

There are still a considerable number of coevents that are not 2-generated by any $q$-measures. For example, we conjecture that $\omega _1^*\oplus\omega _2^*\oplus\omega _3^*$ is not 2-generated by a $q$-measure. For this reason we introduce what we believe (but have not yet proved) is a more general condition. A $q$-measure $\mu$ on
$\ascript$ \textit{actualizes} $\phi\in\ascript ^*$ if there exists a strictly positive symmetric function
$f\colon\Omega\times\Omega\to\real$ such that
\begin{equation*}
\mu (A)=\int\sqbrac{\int _Af(\omega ,\omega ')d\phi (\omega )}d\phi (\omega ')
\end{equation*}
for all $A\in\ascript$. Although we have not been able to show that actualization is more general than 1-generation, the next result shows that this is usually the case.

\begin{thm}       
\label{thm31}
If $\mu$ 1-generates $\phi$ and $\mu (\Omega )\ne 0$, then $\mu$ actualizes $\phi$.
\end{thm}
\begin{proof}
Since $\mu$ 1-generates $\phi$, there exists a density $f$ such that $\mu (A)=\int _Afd\phi$ for all $A\in\ascript$. Define the strictly positive symmetric function $g\colon\Omega\times\Omega\to\real$ by $g(\omega ,\omega ')=f(\omega )f(\omega ')/\mu (\Omega )$. We then have that
\begin{align*}
\int\sqbrac{\int _Ag(\omega ,\omega ')d\phi (\omega )}d\phi (\omega ')
  &=\frac{1}{\mu (\Omega )}\int\sqbrac{\int _Af(\omega )f(\omega ')d\phi (\omega )}d\phi (\omega ')\\
  &=\frac{1}{\mu (\Omega )}\int f(\omega ')\mu (A)d\phi (\omega ')=\mu (A)
\end{align*}
for all $A\in\ascript$. Hence, $\mu$ actualizes $\phi$ with density $g$.
\end{proof}

\section{Actualization} 
The 1- and 2-generation filters have already been considered in \cite{gud093}. Since we have just introduced the actualization filter in this paper, we shall now discuss it in more detail. Let $\Omega _2=\brac{\omega _1,\omega _2}$, 
$\ascript _2=2^{\Omega _2}$ and let $\ascript _2^*$ be the corresponding full anhomomorphic logic. We shall show that every coevent in $\ascript _2^*$ is actualized and shall characterize the actualizing $q$-measures. Now
$\ascript _2^*$ has the eight elements
$0$, $\omega _1^*$, $\omega _2^*$, $\omega _1^*\oplus\omega _2^*$, $\omega _1^*\omega _2^*$,
$\omega _1^*\oplus\omega _1^*\omega _2^*$, $\omega _2^*\oplus\omega _1^*\omega _2^*$ and
$1=\omega _1^*\oplus\omega _2^*\oplus\omega _1^*\omega _2^*$. It is clear that $0$ is actualized by the zero
$q$-measure. For $a>0$ we define the
\textit{Dirac measure} $a\delta _{\omega _1}$ by
\begin{equation*}
a\delta _{\omega _1}(A)=
\begin{cases}
 a&\text{if }\omega _1\in A\\
 0&\text{if }\omega _1\notin A
\end{cases}
\end{equation*}
Since
\begin{equation*}
a\delta _{\omega _1}(A)=\int\sqbrac{\int _Aad\omega _1^*(\omega )}d\omega _1^*(\omega ')
\end{equation*}
it follows that $a\delta _{\omega _1}$ are the only $q$-measures that actualize $\omega _1^*$ and a similar result holds for $\omega _2^*$.

\begin{lem}       
\label{lem41}
A $q$-measure $\mu$ on $\ascript _2$ actualizes $\omega _1^*\oplus\omega _2^*$ if and only if
\begin{equation}         
\label{eq41}
\mu (\Omega _2)=\max\paren{\mu (\omega _1),\mu (\omega _2)}-\min\paren{\mu (\omega _1),\mu (\omega _2)}
\end{equation}
\end{lem}
\begin{proof}
Let $\phi = \omega _1^*\oplus\omega _2^*$ and let $f\colon\Omega _2\times\Omega _2\to\real$ be a strictly positive symmetric function satisfying
\begin{equation}         
\label{eq42}
f(\omega _1,\omega _2)\le f(\omega _1,\omega _1)\le f(\omega _2,\omega _2)
\end{equation}
For $A\in\ascript _2$ define $g_A\colon\Omega _2\to\real ^+$ by
\begin{equation}         
\label{eq43}
g_A(\omega ')=\int _Af(\omega ,\omega ')d\phi (\omega )
\end{equation}
We then have
\begin{align*}
g_{\brac{\omega _1}}(\omega _1)&=\int _{\brac{\omega _1}}f(\omega ,\omega _1)d\phi (\omega )
  =f(\omega _1,\omega _1)\\
g_{\brac{\omega _1}}(\omega _2)&=\int _{\brac{\omega _1}}f(\omega ,\omega _2)d\phi (\omega )
  =f(\omega _1,\omega _2)
\end{align*}
Hence, if $\mu$ actualizes $\phi$ with $\phi$-density $f$ then
\begin{equation*}
\mu (\omega _1)=\int g_{\brac{\omega _1}}(\omega ')d\phi (\omega ')
  =f(\omega _1,\omega _1)-f(\omega _1,\omega _2)
\end{equation*}
and similarly, $\mu (\omega _2)=f(\omega _2,\omega _2)-f(\omega _1,\omega _2)$. We also have that
\begin{align*}
g_{\Omega _2}(\omega _1)&=\int f(\omega ,\omega _1)d\phi (\omega )=f(\omega _1,\omega _1)
  -f(\omega _1,\omega _2)\\
g_{\Omega _2}(\omega _2)&=\int f(\omega ,\omega _2)d\phi (\omega )=f(\omega _2,\omega _2)
  -f(\omega _1,\omega _2)
\end{align*}
Hence,
\begin{align*}
\mu (\Omega _2)&=\int g_{\Omega _2}(\omega ')d\phi (\omega ')=f(\omega _2,\omega _2)
  -f(\omega _1,\omega _1)=\mu (\omega _2)-\mu (\omega _1)\\
  &=\max\paren{\mu (\omega _1),\mu (\omega _2)}-\min\paren{\mu (\omega _1),\mu (\omega _2)}
\end{align*}
The ``only if'' statement of the theorem holds because we obtain similar results for all orderings of
$f(\omega _1,\omega _1)$, $f(\omega _2,\omega _2)$, $f(\omega _1,\omega _2)$.

Conversely, suppose that \eqref{eq41} holds. We can assume without loss of generality that
$\mu (\omega _1)\le\mu (\omega _2)$ so that $\mu (\Omega _2)=\mu (\omega _2)-\mu (\omega _1)$. Let
$f\colon\Omega _2\times\Omega _2\to\real$ be the strictly positive symmetric function given by
$f(\omega _1,\omega _2)=f(\omega _2,\omega _1)=1$, $f(\omega _i,\omega _i)=\mu (\omega _i)+1$, $i=1,2$. Then \eqref{eq42} holds so by our previous work we have for $i=1,2$ that
\begin{align*}
\int\sqbrac{\int _{\brac{\omega _i}}f(\omega ,\omega ')d\phi (\omega )}d\phi (\omega ')
  &=f(\omega _i,\omega _i)-f(\omega _1,\omega _2)=\mu (\omega _i)\\
\intertext{and}
\int\sqbrac{\int f(\omega ,\omega ')d\phi (\omega )}d\phi (\omega ')
  &=f(\omega _2,\omega _2)-f(\omega _1,\omega _1)\\
  &=\mu (\omega _2)-\mu (\omega _1)=\mu (\Omega _2)
\end{align*}
Hence, $\mu$ actualizes $\phi$ with $\phi$-density $f$.
\end{proof}

Notice if we let $\mu (\omega _2)=0$ and $\mu (\omega _1)=\mu (\Omega _2)=1$ in Lemma~\ref{lem41}, $\mu$ becomes the Dirac measure $\delta _{\omega _1}$. This shows that $\delta _{\omega _1}$ actualizes both
$\omega _1^*$ and $\omega _1^*\oplus\omega _2^*$ so actualizing $q$-measures need not be unique. Also, note that $\omega _1^*\oplus\omega _2^*$ is not $\delta _{\omega _1}$-preclusive.

\begin{lem}       
\label{lem42}
A $q$-measure $\mu$ on $\ascript _2$ actualizes $\omega _1^*\omega _2^*$ if and only if
$\mu (\omega _1)=\mu (\omega _2)=0$ and $\mu (\Omega _2)>0$
\end{lem}
\begin{proof}
Let $\phi =\omega _1^*\omega _2^*$ and let $f\colon\Omega _2\times\Omega _2\to\real$ be a strictly positive symmetric function satisfying
\begin{equation*}
f(\omega _1,\omega _2)\le f(\omega _1,\omega _1)\le f(\omega _2,\omega _2)
\end{equation*}
Employing the notation of \eqref{eq43} we have that
\begin{equation*}
g_{\brac{\omega _1}}(\omega _1)=\int _{\brac{\omega _1}}f(\omega ,\omega _1)d\phi (\omega )=0
\end{equation*}
and similarly, $g_{\brac{\omega _1}}(\omega _2)=0$. Hence, if $\mu$ actualizes $\phi$ with $\phi$-density $f$ then
\begin{equation*}
\mu (\omega _1)=\int g_{\brac{\omega _1}}(\omega ')d\phi (\omega ')=0
\end{equation*}
and similarly, $\mu (\omega _2)=0$. We also have that
\begin{align*}
g_{\Omega _2}(\omega _1)&=\int f(\omega ,\omega _1)d\phi (\omega )=f(\omega _1,\omega _2)\\
g_{\Omega _2}(\omega _2)&=\int f(\omega ,\omega _2)d\phi (\omega )=f(\omega _1,\omega _2)
\end{align*}
Hence,
\begin{equation*}
\mu (\Omega _2)=\int g_{\Omega _2}(\omega ')d\phi (\omega ')=f(\omega _1,\omega _2)
\end{equation*}
The ``only if'' statement of the theorem holds because we obtain similar results for all orderings of
$f(\omega _1,\omega _1)$, $f(\omega _2,\omega _2)$, $f(\omega _1,\omega _2)$.

Conversely, suppose that $\mu (\omega _1)=\mu (\omega _2)=0$ and $\mu (\Omega _2)>0$. Let
$f\colon\Omega _2\times\Omega _2\to\real$ be the strictly positive symmetric function given by
$f(\omega _i,\omega _j)=\mu (\Omega _2)$, $i=1,2$. By our previous work $\mu$ actualizes $\phi$ with
$\phi$-density $f$.
\end{proof}

Notice that the $q$-measure in Lemma~\ref{lem42} is not regular.

\begin{lem}       
\label{lem43}
A $q$-measure $\mu$ on $\ascript _2$ actualizes 
$1=\omega _1^*\oplus\omega _2^*\oplus\omega _1^*\omega _2^*$ if and only if
$\mu (\omega _1),\mu (\omega _2)\ne 0$ and $\mu (\Omega _2)=\max\paren{\mu (\omega _1),\mu (\omega _2)}$.
\end{lem}
\begin{proof}
Let $\phi =1$ and let $f\colon\Omega _2\times\Omega _2\to\real$ be a strictly positive symmetric function satisfying
\begin{equation*}
f(\omega _1,\omega _2)\le f(\omega _1,\omega _1)\le f(\omega _2,\omega _2)
\end{equation*}
Employing the notation of \eqref{eq43} we have that
\begin{align*}
g_{\brac{\omega _1}}(\omega _1)&=\int _{\brac{\omega _1}}f(\omega ,\omega _1)d\phi (\omega )
  =f(\omega _1,\omega _1)\\
g_{\brac{\omega _1}}(\omega _2)&=\int _{\brac{\omega _1}}f(\omega ,\omega _2)d\phi (\omega )
  =f(\omega _1,\omega _2)
\end{align*}
Hence, if $\mu$ actualizes $\phi$ with $\phi$-density $f$ then
\begin{equation*}
\mu (\omega _1)=\int g_{\brac{\omega _1}}(\omega ')d\phi (\omega ')=f(\omega _1,\omega _1)
\end{equation*}
and similarly, $\mu (\omega _2)=f(\omega _2,\omega _2)$. We also have that
\begin{align*}
g_{\Omega _2}(\omega _1)&=\int f(\omega ,\omega _1)d\phi (\omega )=f(\omega _1,\omega _1)\\
g_{\Omega _2}(\omega _2)&=\int f(\omega ,\omega _2)d\phi (\omega )=f(\omega _2,\omega _2)
\end{align*}
Hence,
\begin{align*}
\mu (\Omega _2)&=\int g_{\Omega _2}(\omega ')d\phi (\omega ')=f(\omega _2,\omega _2)=\mu (\omega _2)\\
&=\max\paren{\mu (\omega _1),\mu (\omega _2)}
\end{align*}
The ``only if'' statement of the theorem holds because we obtain similar results for all orderings of
$f(\omega _1,\omega _1)$, $f(\omega _2,\omega _2)$, $f(\omega _1,\omega _2)$.

Conversely, suppose that $\mu (\omega _1),\mu (\omega _2)\ne 0$ and
$\mu (\Omega _2)=\max\paren{\mu (\omega _1),\mu (\omega _2)}$. We can assume without loss of generality that
$\mu (\omega _1)\le\mu (\omega _2)$ so that $\mu (\Omega_2)=\mu (\omega _2)$. Define the strictly positive symmetric function $f\colon\Omega _2\times\Omega _2\to\real$ by $f(\omega _2,\omega _2)=\mu (\omega _2)$ and
\begin{equation*}
f(\omega _1,\omega _2)=f(\omega _2,\omega _1)=f(\omega _1,\omega _1)=\mu (\omega _1)
\end{equation*}
By our previous work, $\mu$ actualizes $\phi$ with $\phi$-density $f$.
\end{proof}

\begin{lem}       
\label{lem44}
A $q$-measure $\mu$ on $\ascript _2$ actualizes $\omega _1^*\oplus\omega _1^*\omega _2^*$ if and only if
$\mu (\omega _2)=0$.
\end{lem}
\begin{proof}
Let $\phi =\omega _1^*\oplus\omega _1^*\omega _2^*$. If $\mu$ actualizes $\phi$ then $\mu (\omega _2)=0$ because $\phi (\omega _2)=0$. Conversely, suppose $\mu (\omega _2)=0$. Let
$f\colon\Omega _2\times\Omega _2\to\real$ be the strictly positive symmetric function given by
$f(\omega _2,\omega _2)=1$,
\begin{align*}
f(\omega _1,\omega _2)&=f(\omega _2,\omega _1)=1+\mu (\Omega _2)+\mu (\omega _1)\\
f(\omega _1,\omega _1)&=1+\mu (\Omega _2)+2\mu (\omega _1)
\end{align*}
Since $g_{\brac{\omega _1}}(\omega _1)=f(\omega _1,\omega _1)$,
$g_{\brac{\omega _1}}(\omega _2)=f(\omega _1,\omega _2)$ we have that
\begin{equation*}
\int g_{\brac{\omega _2}}(\omega ')d\phi (\omega ')=f(\omega _1,\omega _1)-f(\omega _1,\omega _2)
=\mu (\omega _1)
\end{equation*}
Of course, $g_{\brac{\omega _2}}=0$ so that
\begin{equation*}
\int g_{\brac{\omega _2}}(\omega ')d\phi (\omega ')=0=\mu (\omega _2)
\end{equation*}
We also have
\begin{align*}
g_{\Omega _2}(\omega _2)&=\int f(\omega ,\omega _1)d\phi (\omega )
  =f(\omega _1,\omega _1)-f(\omega _1,\omega _2)=\mu (\omega _1)\\
  g_{\Omega _2}(\omega _2)&=\int f(\omega ,\omega _2)d\phi (\omega )
  =f(\omega _1,\omega _2)-f(\omega _2,\omega _2)=\mu (\Omega _2)+\mu (\omega _1)
\end{align*}
Hence,
\begin{equation*}
\int g_{\Omega _2}(\omega ')d\phi (\omega ')=g_{\Omega _2}(\omega _2)-g_{\Omega _2}(\omega _1)
=\mu (\Omega _2)
\end{equation*}
We conclude that $\mu$ actualizes $\phi$ with $\phi$-density $f$.
\end{proof}

These lemmas show that the actualization filter need not produce a unique coevent. That is, a $q$-measure may actualize more than one coevent. For example let $f$ be the density function given by $f(\omega _1,\omega _1)=2$ and
\begin{equation*}
f(\omega _1,\omega _2)=f(\omega _2,\omega _1)=f(\omega _2,\omega _2)=1
\end{equation*}
Then the Dirac measure $\delta _{\omega _1}$ actualizes both $\omega _1^*\oplus\omega _2^*$ and
$\omega _1^*\oplus\omega _1^*\omega _2^*$ with density $f$. However, $\omega _1^*\oplus\omega _2^*$ is not
$\delta _{\omega _1}$-preclusive so a preclusivity filter would eliminate $\omega _1^*\oplus\omega _2^*$. Unfortunately, $\delta _{\omega _1}$ also actualizes $\omega _1^*$ with density $g(\omega _i,\omega _j)=1$, $i,j=1,2$. Of course, all these coevents are quadratic. By contrast, we shall show in Section~5 that if a quadratic coevent is 1- or 2-generated by a $q$-measure, then it is unique.

Let $\Omega _3=\brac{\omega _1,\omega _2,\omega _3}$, $\ascript _3=2^{\Omega _3}$ and let $\ascript _3^*$ be the corresponding full anhomomorphic logic. Since $\ab{\ascript _3^*}=2^7=128$ we cannot discuss them all so we consider a few examples.

\begin{exam}{1}    
We have shown in \cite{gud093} that $\phi =\omega _1^*\oplus\omega _2^*\oplus\omega _3^*$ is not 1-generated and we conjecture that $\phi$ is also not 2-generated. We now show that $\phi$ is actualized by the $q$-measure
$\mu$ on $\ascript _3$ given by $\mu (\emptyset )=0$, $\mu (\omega _1)=5$,
$\mu (\omega _2)=\mu\paren{\brac{\omega _2,\omega _3}}=3$,
$\mu (\omega _3)=\mu\paren{\brac{\omega _1,\omega _2}}=6$, $\mu\paren{\brac{\omega _1,\omega _3}}=9$ and
$\mu (\Omega _3)=4$. To show that $\mu$ is indeed a $q$-measure we have that 
\begin{equation*}
\sum _{i<j=1}^3\mu\paren{\brac{\omega _i,\omega _j}}-\sum _{i=1}^3\mu (\omega _i)
  =6+9+3-5-3-6=4=\mu (\Omega _3)
\end{equation*}
Define the density function $f$ by $f(\omega _1,\omega _2)=f(\omega _1,\omega _3)=1$,
$f(\omega _2,\omega _3)=f(\omega _1,\omega _1)=5$, $f(\omega _2,\omega _2)=7$, $f(\omega _3,\omega _3)=10$. For all $A\in\ascript _3$ let
\begin{equation}         
\label{eq44}
\mu '(A)=\int\sqbrac{\int _Af(\omega ,\omega ')d\phi (\omega )}d\phi (\omega ')
\end{equation}
Employing the notation \eqref{eq43} we obtain
\begin{align*}
g_{\brac{\omega _1}}(\omega _1)&=f(\omega _1,\omega _1)=5,\quad
g_{\brac{\omega _1}}(\omega _2)=f(\omega _1,\omega _2)=1,\\
g_{\brac{\omega _1}}(\omega _3)&=f(\omega _1,\omega _3)=1
\end{align*}
It follows that
\begin{equation*}
\mu '(\omega _1)=f(\omega _1,\omega _2)+f(\omega _1,\omega _1)-f(\omega _1,\omega _3)=5
\end{equation*}
In a similar way we have that
\begin{align*}
\mu '(\omega _2)&=f(\omega _1,\omega _2)+f(\omega _2,\omega _2)-f(\omega _2,\omega _3)=3\\
\mu '(\omega _3)&=f(\omega _1,\omega _3)+f(\omega _3,\omega _3)-f(\omega _2,\omega _3)=6
\end{align*}
We also have that
\begin{align*}
g_{\brac{\omega _1,\omega _2}}(\omega _1)&=f(\omega _1,\omega _1)-f(\omega _1,\omega _2)=4\\
g_{\brac{\omega _1,\omega _2}}(\omega _2)&=f(\omega _2,\omega _2)-f(\omega _1,\omega _2)=6\\
g_{\brac{\omega _1,\omega _2}}(\omega _3)&=f(\omega _2,\omega _3)-f(\omega _1,\omega _3)=4
\end{align*}
and hence, $\mu '\paren{\brac{\omega _1,\omega _2}}=6$. Continuing the computations gives
\begin{align*}
g_{\brac{\omega _1,\omega _3}}(\omega _1)&=f(\omega _1,\omega _1)-f(\omega _1,\omega _3)=4\\
g_{\brac{\omega _1,\omega _3}}(\omega _2)&=f(\omega _2,\omega _3)-f(\omega _1,\omega _2)=4\\
g_{\brac{\omega _1,\omega _3}}(\omega _3)&=f(\omega _3,\omega _3)-f(\omega _1,\omega _3)=9
\end{align*}
and hence, $\mu '\paren{\brac{\omega _1,\omega _3}}=9$. We also have that
\begin{align*}
g_{\brac{\omega _2,\omega _3}}(\omega _1)&=f(\omega _1,\omega _3)-f(\omega _1,\omega _2)=0\\
g_{\brac{\omega _2,\omega _3}}(\omega _2)&=f(\omega _2,\omega _2)-f(\omega _2,\omega _3)=2\\
g_{\brac{\omega _2,\omega _3}}(\omega _3)&=f(\omega _3,\omega _3)-f(\omega _2,\omega _3)=5
\end{align*}
and hence, $\mu '\paren{\brac{\omega _2,\omega _3}}=3$. Finally,
\begin{align*}
g_{\Omega _3}(\omega _1)&=f(\omega _1,\omega _2)+f(\omega _1,\omega _1)
  -f(\omega _1,\omega _3)=5\\
g_{\Omega _3}(\omega _2)&=f(\omega _1,\omega _2)+f(\omega _2,\omega _2)
  -f(\omega _2,\omega _3)=3\\
g_{\Omega _3}(\omega _3)&=f(\omega _1,\omega _2)+f(\omega _3,\omega _3)
  -f(\omega _2,\omega _3)=6
\end{align*}
and hence, $\mu '(\Omega _3)=4$. Since $\mu (A)=\mu '(A)$ for all $A\in\ascript _3$ we conclude that $\mu$ actualizes $\phi$.\hfill\qedsymbol
\end{exam} 

\begin{exam}{2}    
This example shows that $\delta _{\omega _1}$ actualizes the coevent 
$\phi=\omega _1^*\oplus\omega _1^*\omega _2^*\omega _3^*$ with density given by $f(\omega _i,\omega _j)=1$
for $i\ne j=1,2,3$, $f(\omega _i,\omega _i)=2$, $i=1,2,3$. As before, define $\mu '$ by \eqref{eq45}. As in previous calculations, we have $g_{\brac{\omega _1}}(\omega _i)=f(\omega _1,\omega _i)$, $i=1,2,3$ and hence,
\begin{equation*}
\mu '(\omega _1)=f(\omega _1,\omega _1)-f(\omega _1,\omega _3)=1
\end{equation*}
It is clear that
\begin{equation*}
g_{\brac{\omega _2}}(\omega _i)=g_{\brac{\omega _3}}(\omega _i)=0
\end{equation*}
for $i=1,2,3$ and hence, $\mu '(\omega _2)=\mu '(\omega _3)=0$. We also have that
\begin{equation*}
g_{\brac{\omega _1,\omega _2}}(\omega _i)=g_{\brac{\omega _1,\omega _3}}(\omega _i)=f(\omega _1,\omega _i)
\end{equation*}
for $i=1,2,3$ and hence,
\begin{equation*}
\mu '\paren{\brac{\omega _1,\omega _2}}=\mu '\paren{\brac{\omega _1\omega _3}}
  =f(\omega _1,\omega _1)-f(\omega _1,\omega _2)=1
\end{equation*}
Moreover, $g_{\brac{\omega _2,\omega _3}}(\omega _i)=0$ for $i=1,2,3$ so that
$\mu '\paren{\brac{\omega _2,\omega _3}}=0$. Finally,
$g_{\Omega _3}(\omega _1)=f(\omega _1,\omega _1)-f(\omega _1,\omega _2)=1$,
$g_{\Omega _3}(\omega _2)=g_{\Omega _3}(\omega _3)=0$ so that $\mu '(\Omega _3)=1$. Since
$\delta _{\omega _1}(A)=\mu '(A)$ for all $A\in\ascript _3$ we conclude that $\delta _{\omega _1}$ actualizes $\phi$. Of course, $\delta _{\omega _1}$ also actualizes $\omega _1^*$ so we again have nonuniqueness.
\hfill\qedsymbol
\end{exam} 

\begin{exam}{3}    
Calculations similar to those in the previous two examples show that
$\phi =\omega _1^*\oplus\omega _2^*\oplus\omega _3^*\oplus\omega _1^*\omega _2^*$ is actualized by the
$q$-measure $\mu$ given by $\mu (\emptyset )=0$, $\mu (\omega _1)=\mu (\Omega _3)=1$,
$\mu (\omega _2)=\mu\paren{\brac{\omega _1,\omega _2}}=4$,
$\mu (\omega _3)=\mu\paren{\brac{\omega _1,\omega _3}}=3$ and $\mu\paren{\brac{\omega _2,\omega _3}}=2$. The corresponding density is given by
\begin{align*}
f(\omega _1,\omega _1)&=f(\omega _2,\omega _2)=2\\
f(\omega _3,\omega _3)&=f(\omega _1,\omega _2)=4\\
f(\omega _1,\omega _3)&=5,\ f(\omega _2,\omega _3)=8
\end{align*}
It can be shown that $\phi$ is also actualized by the $q$-measure $\nu$ given by $\nu (\emptyset )=0$,
\begin{align*}
\nu (\omega _1)&=\nu\paren{\brac{\omega _2,\omega _3}}=\nu (\Omega _3)=0\\
\nu (\omega _2)&=\nu (\omega _3)=\nu\paren{\brac{\omega _1,\omega _2}}
  =\nu\paren{\brac{\omega _1,\omega _3}}=1
\end{align*}
The corresponding density is given by $f(\omega _i,\omega _j)=1$ for $i,j=1,2,3$, $(i,j)\ne (2,3)$ or $(3,2)$ and $f(\omega _2,\omega _3)=2$. In the second case, $\phi$ is not $\nu$-preclusive.\hfill\qedsymbol
\end{exam}

\section{Generation} 
This section discusses existence and uniqueness properties of 1- and 2-\newline
generated coevents. We first consider existence. It is clear that any coevent in $\ascript _2^*$ is 1- and 2-generated (and actualized). We now discuss coevents in $\ascript ^*=\ascript _n^*$ that are 1-generated.

A $q$-measure on $\ascript$ whose only values are $0$ or $1$ is called a \textit{pure} $q$-\textit{measure}. A pure $q$-measure can also be thought of as a coevent in $\ascript ^*$ and such coevents are called
\textit{pure coevents}. Thus, a pure coevent is an element of $\ascript ^*$ that is also a $q$-measure. Although this appears to be rather specialized, there are quite a few pure coevents and most $q$-measures can be written as convex combinations of pure $q$-measures.

It is clear that any $\phi\in\ascript _2^*$ is a pure coevent. It can be shown that of the 128 coevents in
$\ascript _3^*$, 34 are pure \cite{gud092}.

\begin{exam}{4}    
Examples of pure coevents in $\ascript _3^*$ are $\omega _1^*$, $\omega _1^*\oplus\omega _2^*$,
$\omega _1^*\oplus\omega _1^*\omega _2^*$, $\omega _1^*\omega _2^*$,
$\omega _1^*\oplus\omega _2^*\oplus\omega _1^*\omega _2^*$,
$\omega _1^*\oplus\omega_1^*\omega _2^*\oplus\omega _2^*\omega _3^*$,
$\omega _1^*\oplus\omega _2^*\oplus\omega _1^*\omega _2^*\oplus\omega _1\omega _3^*$,
$\omega _1^*\oplus\omega _1^*\omega _2^*\oplus\omega _1^*\omega _3^*\oplus\omega _2^*\omega _3^*$,
$\omega _1^*\oplus\omega _2^*\oplus\omega _3^*\oplus\omega _1^*\omega _2^*
  \oplus\omega _1^*\omega _3^*\oplus\omega _2^*\omega _3^*$
and the rest are obtained by symmetry. An example of a $\phi\in\ascript _3^*$ that is not pure is
$\phi =\omega _1^*\oplus\omega _2^*\oplus\omega _3^*$. Indeed, $\phi (\Omega _3)=1$ and
\begin{equation*}
\phi\paren{\brac{\omega _1,\omega _2}}+\phi\paren{\brac{\omega _1,\omega _3}}
+\phi\paren{\brac{\omega _2,\omega _3}}-\phi (\omega _1)-\phi (\omega _2)-\phi (\omega _3)=-3
\end{equation*}
Another example of a nonpure element of $\ascript _3^*$ is $\psi =\omega _1^*\oplus\omega _2^*\omega _3^*$.
Indeed, $\psi (\Omega _3)=0$ and
\begin{equation*}
\psi\paren{\brac{\omega _1,\omega _2}}+\psi\paren{\brac{\omega _1,\omega _3}}
+\psi\paren{\brac{\omega _2,\omega _3}}-\psi (\omega _1)-\psi (\omega _2)-\psi (\omega _3)=2\quad\qedsymbol
\end{equation*}
\end{exam}
\medskip

\begin{lem}       
\label{lem51} If $\phi\in\ascript ^*$ is pure, then $\phi$ is quadratic.
\end{lem}
\begin{proof}
We must show that if $\phi$ satisfies
\begin{equation}         
\label{eq51}
\phi (A\cupdot B\cupdot C)=\phi (A\cupdot B)+\phi (A\cupdot C)+\phi (B\cupdot C)-\phi (A)-\phi (B)-\phi (C)
\end{equation}
then $\phi$ satisfies
\begin{equation}         
\label{eq52}
\phi (A\cupdot B\cupdot C)
  =\phi (A\cupdot B)\oplus\phi (A\cupdot C)\oplus\phi (B\cupdot C)\oplus\phi (A)\oplus\phi (B)\oplus\phi (C)
\end{equation}

Suppose the left hand side of \eqref{eq52} is $1$. Then there are an odd number of $1$s on the right hand side of
\eqref{eq51}. Hence, the right hand side of \eqref{eq52} is $1$. Suppose the left hand side of \eqref{eq52} is $0$. Then there are an even number of $1$s on the right hand side of \eqref{eq51}. Hence, the right hand side of
\eqref{eq52} is $0$. We conclude that \eqref{eq52} holds so $\phi$ is quadratic.
\end{proof}

The converse Lemma~\ref{lem51} does not hold. For instance, in Example~4 we showed that the quadratic coevent
$\omega _1^*\oplus\omega _2^*\oplus\omega _3^*$ is not pure.

\begin{thm}       
\label{thm52}
A coevent $\phi\in\ascript ^*$ is 1-generated if and only if $\phi$ is pure.
\end{thm}
\begin{proof}
It is clear that if $\phi$ is pure, then $\phi$ 1-generates itself. Conversely, suppose
$\phi\in\ascript _n^*$ is 1-generated by the $q$-measure $\mu$ with $\phi$-density $f$. We can reorder the
$\omega _i\in\Omega _n$ if necessary and assume that $f(\omega _i)=a_i$, $i=1,\ldots ,n$, where
$0<a_1\le a_2\le\cdots\le a_n$. Since $\mu (A)=\int _Afd\phi$ for every $A\in\ascript _n$ we have
$\mu (\omega _i)=a_i\phi (\omega _i)$, $i=1,2,\ldots ,n$ and for $i<j=1,2,\ldots ,n$ that 
\begin{equation}         
\label{eq53}
\mu\paren{\brac{\omega _i,\omega _j}}=\int _{\brac{\omega _i,\omega _j}}fd\phi
  =a_i\phi\paren{\brac{\omega _i,\omega _j}}+(a_j-a_i)\phi (\omega _j)
\end{equation}
Letting $i<j<k$ with $i,j,k=1,\ldots ,n$, and $A=\brac{\omega _i,\omega _j,\omega _k}$ we have
\begin{equation}         
\label{eq54}
\mu (A)=\int _Afd\phi =a_i\phi(A)+(a_j-a_i)\phi\paren{\brac{\omega _j,\omega _k}}+(a_k-a_j)\phi (\omega _k)
\end{equation}
Applying \eqref{eq53} and grade-2 additivity of $\mu$ gives
\begin{align}         
\label{eq55}
\mu (A)&=a_i\phi\paren{\brac{\omega _i,\omega _j}}+(a_j-a_i)\phi (\omega _j)
  +a_i\phi\paren{\brac{\omega _i,\omega _k}}+(a_k-a_i)\phi (\omega _k)\notag\\
  &\quad +a_j\phi\paren{\brac{\omega _j,\omega _k}}+(a_k-a_j)\phi (\omega _k)-a_i\phi (\omega _i)
  -a_j\phi (\omega _j)-a_k\phi (\omega _k)
\end{align}
Equating \eqref{eq54} and \eqref{eq55} we see that all the terms cancel except those with a factor of $a_i$. Canceling the $a_i$ gives
\begin{equation}         
\label{eq56}
\phi (A)=\phi\paren{\brac{\omega _i,\omega _j}}+\phi\paren{\brac{\omega _i,\omega _k}}
  +\phi\paren{\brac{\omega _j,\omega _k}}-\phi (\omega _i)-\phi (\omega _j)-\phi (\omega _k)
\end{equation}

We can now proceed by induction to show that $\phi$ satisfies \eqref{eq32} and thus is a $q$-measure. Instead of carrying out the general induction step which is quite cumbersome, we shall verify \eqref{eq32} for
$B=\brac{\omega _i,\omega _j,\omega _k,\omega _l}$ where $i,j,k,l\in\brac{1,2,\ldots n}$ with $i<j<k<l$. As in
\eqref{eq53} we have by \eqref{eq56} that
\begin{align}         
\label{eq57}
\mu (B)&=\int _Bfd\phi =a_i\phi (B)+(a_j-a_i)\phi\paren{\brac{\omega _j,\omega _k,\omega _l}}\notag\\
  &\qquad +(a_k-a_j)\phi\paren{\brac{\omega _k,\omega _l}}+(a_l-a_k)\phi (\omega _l)\\
  &=a_i\phi (B)+(a_j-a_i)\left [\phi\paren{\brac{\omega _j,\omega _k}}+\phi\paren{\brac{\omega _j,\omega _l}}
  +\phi\paren{\brac{\omega _k,\omega _l}}\right.\notag\\
  &\quad \left.- \phi (\omega _j)-\phi (\omega _k)-\phi (\omega _l)\right ]
  +(a_k-a_j)\phi\paren{\brac{\omega _k,\omega _l}}+(a_l-a_k)\phi (\omega _l)\notag
\end{align}
Again, applying \eqref{eq53} and grade-2 additivity of $\mu$ gives
\begin{align}         
\label{eq58}
\mu (B)&=a_i\phi\paren{\brac{\omega _i,\omega _j}}+(a_j-a_i)\phi (\omega _j)
  +a_i\phi\paren{\brac{\omega _i,\omega _k}}+(a_k-a_i)\phi (\omega _k)\notag\\
  &\quad +a_i\phi\paren{\brac{\omega _i,\omega _l}}+(a_l-a_i)\phi (\omega _l)
  +a_j\phi\paren{\brac{\omega _j,\omega _k}}+(a_k-a_j)\phi (\omega _k)\notag\\
  &\quad +a_j\phi\paren{\brac{\omega _j,\omega _l}}+(a_l-a_j)\phi (\omega _l)
  +a_k\phi\paren{\brac{\omega _k,\omega _l}}+(a_l-a_k)\phi (\omega _l)\notag\\
  &\quad -2\sqbrac{a_i\phi (\omega _i)+a_j\phi (\omega _j)+a_k\phi (\omega _k)+a_l\phi (\omega _l)}
\end{align}
Equating \eqref{eq57} and \eqref{eq58} we see that all the terms cancel except those with a factor $a_i$. Canceling the $a_i$ shows that \eqref{eq32} holds for $\phi$ with $m=4$.
\end{proof}

It follows from Theorem~\ref{thm52} that if $\phi$ is 1-generated then $\phi$ is 2-generated. Indeed if $\phi$ is 1-generated then $\phi$ is 2-generated by itself because
\begin{equation*}
\int _A\sqbrac{\int _Ad\phi (\omega )}d\phi (\omega ')=\int _A\phi (A)d\phi =\phi (A)^2=\phi (A)
\end{equation*}
Similar to Theorem~\ref{thm31} it also follows from Theorem~\ref{thm52} that if $\phi$ is unital and 1-generated then $\phi$ is actualized. Indeed, we then have
\begin{equation*}
\int\sqbrac{\int _Ad\phi (\omega )}d\phi (\omega ')=\int\phi (A)d\phi =\phi (A)\phi (\Omega )=\phi (A)
\end{equation*}
Using these filters we see that the actual reality corresponding to a pure $q$-measure is itself. The next example shows that 2-generation is strictly more general than 1-generation.

\begin{exam}{5}    
It is easy to check that $\phi =\omega _1^*\oplus\omega _2^*\oplus\omega _3^*\oplus\omega _1^*\omega _2^*$ is not pure so $\phi$ is not 1-generated. Let $\mu$ be the $q$-measure on $\ascript _3$ defined by
$\mu (\emptyset )=\mu (\Omega _3)=0$, $\mu (\omega _3)=\mu\paren{\brac{\omega _1,\omega _2}}=2$,
\begin{equation*}
\mu (\omega _1)=\mu (\omega _2)
  =\mu\paren{\brac{\omega _1,\omega _3}}=\mu\paren{\brac{\omega _2,\omega _3}}=1
\end{equation*}
It can be shown that $\mu$ 2-generates $\phi$ with $\phi$-density $f$ given by
$f(\omega _1,\omega _1)=f(\omega _2,\omega _2)=1$ and
\begin{equation*}
\hskip 5.5pc
f(\omega _3,\omega _3)=f(\omega _1,\omega _2)=f(\omega _1,\omega _3)=f(\omega _2,\omega _3)=2
\hskip 5.5pc \qedsymbol
\end{equation*}
\end{exam}
\medskip

The next two results concern the uniqueness of 1- and 2-generated coevents.

\begin{thm}       
\label{thm53}
If $\mu$ 1-generates $\phi$ and $\psi$ then $\phi =\psi$.
\end{thm}
\begin{proof}
We have that
\begin{equation*}
\mu (A)=\int _Afd\phi =\int _Agd\psi
\end{equation*}
for all $A\in\ascript$ where $f$ is a $\phi$-density and $g$ is a $\psi$-density for $\mu$. Since
\begin{equation*}
\int _{\brac{\omega}}fd\phi =f(\omega )\phi (\omega )
\end{equation*}
we conclude that $f(\omega )\phi (\omega )=g(\omega )\psi (\omega )$ for every $\omega\in\Omega$. Hence,
$\phi (\omega )=1$ if and only if $\psi (\omega )=1$ and in this case
\begin{equation*}
f(\omega )=g(\omega )=\mu (\omega )
\end{equation*}
Thus, $\phi$ and $\psi$ agree on singleton sets and $\mu (\omega )=0$ if and only if $\phi (\omega )=0$. Suppose that $0<\mu (\omega _1)\le\mu (\omega _2)$. Then $f(\omega _1)\le f(\omega _2)$ and
\begin{equation*}
\phi (\omega _1)=\phi (\omega _2)=\psi (\omega _1)=\psi (\omega _2)=1
\end{equation*}
Similarly, $g(\omega _1)\le g(\omega _2)$. Since
\begin{equation*}
\int _{\brac{\omega _1,\omega _2}}fd\phi =\int _{\brac{\omega _1,\omega _2}}gd\psi
\end{equation*}
we have that 
\begin{align}         
\label{eq59}
f(\omega _1)\phi&\paren{\brac{\omega _1,\omega _2}}+\sqbrac{f(\omega _2)-f(\omega _1)}\phi (\omega _2)\notag\\
  &=g(\omega _1)\psi\paren{\brac{\omega _1,\omega _2}}+\sqbrac{g(\omega _2)-g(\omega _1)}\psi (\omega _2)
\end{align}
Hence, $\phi\paren{\brac{\omega _1,\omega _2}}=\psi\paren{\brac{\omega _1,\omega _2}}$. Next suppose that
$\mu (\omega _1)=0$ and $\mu (\omega _2)\ne 0$. Then $\phi (\omega _1)=\psi (\omega _1)=0$ and
$\phi (\omega _2)=\psi (\omega _2)=1$. Again \eqref{eq59} holds. Hence,
$\phi\paren{\brac{\omega _1,\omega _2}}=0$ if and only if $\psi\paren{\brac{\omega _1,\omega _2}}=0$. 
The case $\mu (\omega _1)=\mu (\omega _2)=0$ is similar. We conclude that $\phi$ and $\psi$ agree on doubleton sets. Since $\phi$ and $\psi$ are quadratic by Lemma~\ref{lem51}, and quadratic coevents are determined by their values on singleton and doubleton sets, $\phi$ and $\psi$ coincide.
\end{proof}

\begin{thm}       
\label{thm54}
If $\mu$ 2-generates $\phi$ and $\psi$ and both $\phi$ and $\psi$ are quadratic, then $\phi =\psi$.
\end{thm}
\begin{proof}
We have that
\begin{align}         
\label{eq510}
\mu (A)&=\int _A\sqbrac{\int _Af(\omega ,\omega ')d\phi (\omega )}d\phi (\omega ')\notag\\
 & =\int _A\sqbrac{\int _Ag(\omega ,\omega ')d\psi (\omega )}d\psi (\omega ')
\end{align}
for all $A\in\ascript$. Letting $A=\brac{\omega }$ in \eqref{eq510} we conclude that
\begin{equation*}
f(\omega ,\omega )\phi (\omega )=g(\omega ,\omega )\psi (\omega )
\end{equation*}
for all $\omega\in\Omega$. Hence, $\phi (\omega )=1$ if and only if $\psi (\omega )=1$ and in this case
$f(\omega ,\omega )=g(\omega ,\omega )=\mu (\omega )$. We conclude that $\phi$ and $\psi$ agree on singleton sets and $\mu (\omega )=0$ if and only if $\psi (\omega )=0$. Suppose $0<\mu (\omega _1)\le\mu (\omega _2)$. Then
$f(\omega _1,\omega _1)\le f(\omega _2,\omega _2)$ and
\begin{equation*}
\phi (\omega _1)=\phi (\omega _2)=\psi (\omega _1)=\psi (\omega _2)=1
\end{equation*}
Assume that $f(\omega _2,\omega _2)\le f(\omega _1,\omega _2)$.

\noindent\textbf{Case 1.}\enspace 
$g(\omega _1,\omega _1)\le g(\omega _2,\omega _2)\le g(\omega _1,\omega _2)$\newline
Letting
\begin{equation*}
h_f(\omega ')=\int _{\brac{\omega _1,\omega _2}}f(\omega ,\omega ')d\phi (\omega ),\quad
h_g(\omega ')=\int _{\brac{\omega _1,\omega _2}}g(\omega ,\omega ')d\phi (\omega )
\end{equation*}
we have that
\begin{align*}
h_f(\omega _1)
  &=\mu (\omega _1)\phi\paren{\brac{\omega _1,\omega _2}}+f(\omega _1,\omega _2)-\mu (\omega _1)\\
h_f(\omega _2)
  &=\mu (\omega _2)\phi\paren{\brac{\omega _1,\omega _2}}+f(\omega _1,\omega _2)-\mu (\omega _2)
\end{align*}
Hence,
\begin{align}         
\label{eq511}
\mu\paren{\brac{\omega _1,\omega _2}}
  &=\int _{\brac{\omega _1,\omega _2}}h_f(\omega ')d\phi (\omega ')\notag\\
  &=\sqbrac{\mu (\omega _2)\phi\paren{\brac{\omega _1,\omega _2}}+f(\omega _1,\omega _2)-\mu (\omega _2)}
  \phi\paren{\brac{\omega _1,\omega _2}}\notag\\
  &\quad +\sqbrac{\mu (\omega _1)-\mu (\omega _2)}\phi\paren{\brac{\omega _1,\omega _2}}+\mu (\omega _2)
  -\mu (\omega _1)
\end{align}
If $\phi\paren{\brac{\omega _1,\omega _2}}=0$, then
$\mu\paren{\brac{\omega _1,\omega _2}}=\mu (\omega _2)-\mu (\omega _1)$. Since \eqref{eq511} also applies for $\psi$, we conclude that $\psi\paren{\brac{\omega _1,\omega _2}}=0$. If
$\phi\paren{\brac{\omega _1,\omega _2}}=1$, then
$\mu\paren{\brac{\omega _1,\omega _2}}=f(\omega _1,\omega _2)$. Again, \eqref{eq511} also applies for $\psi$ so $\psi\paren{\brac{\omega _1,\omega _2}}=1$.

\medskip
\noindent\textbf{Case 2.}\enspace  
$g(\omega _1,\omega _1)\le g(\omega _1,\omega _2)\le g(\omega _2,\omega _2)$\newline
We now have that
\begin{align*}
h_g(\omega _1)
  &=\mu (\omega _1)\psi\paren{\brac{\omega _1,\omega _2}}+g(\omega _1,\omega _2)-\mu (\omega _1)\\
h_g(\omega _2)
  &=g(\omega _1,\omega _2)\psi\paren{\brac{\omega _1,\omega _2}}+\mu (\omega _2)-g(\omega _1,\omega _2)
\end{align*}
If $\psi\paren{\brac{\omega _1,\omega _2}}=1$, then
\begin{equation*}
\mu\paren{\brac{\omega _1,\omega _2}}=\int _{\brac{\omega _1,\omega _2}}h_g(\omega ')d\psi (\omega ')
  =\mu (\omega _2)
\end{equation*}
and hence $f(\omega _1,\omega _2)=\mu (\omega _2)$ so that $\phi\paren{\brac{\omega _1,\omega _2}}=1$. If
$\psi\paren{\brac{\omega _1,\omega _2}}=0$, then
\begin{equation*}
\mu\paren{\brac{\omega _1,\omega _2}}=\int _{\brac{\omega _1,\omega _2}}h_g(\omega ')d\psi (\omega ')
  =2g(\omega _1,\omega _2)-\mu (\omega _1)-\mu (\omega _2)
\end{equation*}
or $\mu\paren{\brac{\omega _1,\omega _2}}=\mu (\omega _2)+\mu (\omega _1)-2g(\omega _1,\omega _2)$ whichever is nonnegative. If $\phi\paren{\brac{\omega _1,\omega _2}}=1$, then
\begin{equation*}
2g(\omega _1,\omega _2)-\mu (\omega _1)-\mu (\omega _2)=f(\omega _1,\omega _2)\ge\mu (\omega _2)
\end{equation*}
so that $2g(\omega _1,\omega _2)\ge 2\mu (\omega _2)+\mu (\omega _1)$
which is a contradiction. We could also have
\begin{equation*}
\mu (\omega _2)+\mu (\omega _1)-2g(\omega _1,\omega _2)=f(\omega _1,\omega _2)\ge\mu (\omega _2)
\end{equation*}
so that $2g(\omega _1,\omega _2)\le\mu (\omega _1)$ which is a contradiction. We conclude that
$\psi\paren{\brac{\omega _1,\omega _2}}=0$ if and only if $\phi\paren{\brac{\omega _1,\omega _2}}=0$.

\medskip
\noindent\textbf{Case 3.}\enspace
$g(\omega _1,\omega _2)\le g(\omega _1,\omega _1)\le g(\omega _2,\omega _2)$\newline
We now have that
\begin{align*}
h_g(\omega _1)
  &=g(\omega _1,\omega _2)\psi\paren{\brac{\omega _1,\omega _2}}+\mu (\omega _1)-g(\omega _1,\omega _2)\\
h_g(\omega _2)
  &=g(\omega _1,\omega _2)\psi\paren{\brac{\omega _1,\omega _2}}+\mu (\omega _2)-g(\omega _1,\omega _2)
\end{align*}
Hence,
\begin{align*}
\mu\paren{\brac{\omega _1,\omega _2}}&=\int _{\brac{\omega _1,\omega _2}}h_g(\omega ')d\psi (\omega ')\\
  &=\sqbrac{g(\omega _1,\omega _2)\psi\paren{\brac{\omega _1,\omega _2}}
  +\mu (\omega _1)-g(\omega _1,\omega _2)}\psi\paren{\brac{\omega _1,\omega _2}}\\
  &\quad +\mu (\omega _2)-\mu (\omega _1)
\end{align*}
If $\psi\paren{\brac{\omega _1,\omega _2}}=0$, then
$\mu\paren{\brac{\omega _1,\omega _2}}=\mu (\omega _2)-\mu (\omega _1)$ so that
$\phi\paren{\brac{\omega _1,\omega _2}}=0$. If $\psi\paren{\brac{\omega _1,\omega _2}}=1$, then
\begin{equation*}
\mu\paren{\brac{\omega _1,\omega _2}}=\mu (\omega _2)>\mu (\omega _2)-\mu (\omega _1)
\end{equation*}
This implies that $\phi\paren{\brac{\omega _1,\omega _2}}=1$. There are other cases, including the case where
$f(\omega _1,\omega _2)\le f(\omega _2,\omega _2)$, but the results are similar. It follows that in all possible cases where $\mu (\omega _1),\mu (\omega _2)\ne 0$ we have
$\phi\paren{\brac{\omega _1,\omega _2}}=\psi\paren{\brac{\omega _1,\omega _2}}$.

We now consider the situation in which $\mu (\omega _1)=\mu (\omega _2)=0$.
Then
\begin{equation*}
\phi (\omega _1)=\phi (\omega _2)=\psi (\omega _1)=\psi (\omega _2)=0
\end{equation*}
Assume that
\begin{align}         
\label{eq512}
f(\omega _1,\omega _1)&\le f(\omega _2,\omega _2)\le f(\omega _1,\omega _2)\\
 \intertext{and}
g(\omega _1,\omega _1)&\le g(\omega _2,\omega _2)\le g(\omega _1,\omega _2)\notag
\end{align}
We then have that
\begin{equation*}
h_f(\omega _1)=f(\omega _1,\omega _1)\phi\paren{\brac{\omega _1,\omega _2}},\quad
h_f(\omega _2)=f(\omega _2,\omega _2)\phi\paren{\brac{\omega _1,\omega _2}}
\end{equation*}
Hence,
\begin{equation*}
\mu\paren{\brac{\omega _1,\omega _2}}
  =f(\omega _1,\omega _1)\phi\paren{\brac{\omega _1,\omega _2}}
  =g(\omega _1,\omega _1)\psi\paren{\brac{\omega _1,\omega _2}}
\end{equation*}
and $\phi\paren{\brac{\omega _1,\omega _2}}=\psi\paren{\brac{\omega _1,\omega _2}}$. All the other cases in this situation are similar.

The last situation that needs to be considered is $\mu (\omega _1)=0$, $\mu (\omega _2)>0$. Then
$\phi (\omega _1)=\psi (\omega _1)=0$, $\phi (\omega _2)=\psi (\omega _2)=1$ and
\begin{equation*}
f(\omega _2,\omega _2)=g(\omega _2,\omega _2)=\mu (\omega _2)
\end{equation*}
Assuming that \eqref{eq512} holds, we treat the three cases considered before.

\medskip
\noindent\textbf{Case 1.}\enspace 
$g(\omega _1,\omega _1)\le g(\omega _2,\omega _2)\le g(\omega _1,\omega _2)$\newline
We now have that
\begin{align*}
h_f(\omega _1)&=f(\omega _1,\omega _1)\phi\paren{\brac{\omega _1,\omega _2}}
  +f(\omega _1,\omega _2)-f(\omega _1,\omega _1)\\
h_f(\omega _2)&=\mu (\omega _2)\phi\paren{\brac{\omega _1,\omega _2}}
\end{align*}
Since we obtain similar results for $h_g$, we conclude that
\begin{equation*}
\mu\paren{\brac{\omega _1,\omega _2}}=\mu (\omega _2)\phi\paren{\brac{\omega _1,\omega _2}}
=\mu (\omega _2)\psi\paren{\brac{\omega _1,\omega _2}}
\end{equation*}
Hence, $\phi\paren{\brac{\omega _1,\omega _2}}=\psi\paren{\brac{\omega _1,\omega _2}}$.

\medskip
\noindent\textbf{Case 2.}\enspace  
$g(\omega _1,\omega _1)\le g(\omega _1,\omega _2)\le g(\omega _2,\omega _2)$\newline
As in Case~1 we have that
$\mu\paren{\brac{\omega _1,\omega _2}}=\mu (\omega _2)\phi\paren{\brac{\omega _1,\omega _2}}$. Also,
\begin{align*}
h_g(\omega _1)&=g(\omega _1,\omega _1)\psi\paren{\brac{\omega _1,\omega _2}}
  +g(\omega _1,\omega _2)-g(\omega _1,\omega _1)\\
h_g(\omega _2)&=g(\omega _1,\omega _2)\psi\paren{\brac{\omega _1,\omega _2}}
  +\mu(\omega _2)-g(\omega _1,\omega _2)
\end{align*}
If $\psi\paren{\brac{\omega _1,\omega _2}}=1$, then
\begin{equation*}
\mu\paren{\brac{\omega _1,\omega _2}}=\mu (\omega _2)
  =\mu (\omega _2)\phi\paren{\brac{\omega _1,\omega _2}}
\end{equation*}
so that $\phi\paren{\brac{\omega _1,\omega _2}}=1$. If $\psi\paren{\brac{\omega _1,\omega _2}}=0$, then
\begin{equation*}
\mu\paren{\brac{\omega _1,\omega _2}}=\mu (\omega _2)+g(\omega _1,\omega _1)-2g(\omega _1,\omega _2)
  =\mu (\omega _2)\phi\paren{\brac{\omega _1,\omega _2}}
\end{equation*}
If $\phi\paren{\brac{\omega _1,\omega _2}}=1$ we obtain the contradiction,
$g(\omega _1,\omega _1)=2g(\omega _1,\omega _2)$ so $\phi\paren{\brac{\omega _1,\omega _2}}=0$. Alternatively, we could have $\mu\paren{\brac{\omega _1,\omega _2}}=0$ so again,
$\phi\paren{\brac{\omega _1,\omega _2}}=0$.

\medskip
\noindent\textbf{Case 3.}\enspace
$g(\omega _1,\omega _2)\le g(\omega _1,\omega _1)\le g(\omega _2,\omega _2)$\newline
We now have that
\begin{align*}
\mu\paren{\brac{\omega _1,\omega _2}}&=g(\omega _1,\omega _2)\psi\paren{\brac{\omega _1,\omega _2}}
  +\mu (\omega _2)-g(\omega _1,\omega _2)\\
  &=\mu (\omega _2)\phi\paren{\brac{\omega _1,\omega _2}}
\end{align*}
Just as in Case~2 we conclude that
$\psi\paren{\brac{\omega _1,\omega _2}}=\phi\paren{\brac{\omega _1,\omega _2}}$. The other cases are similar to the three cases considered.

We have shown that $\phi$ and $\psi$ coincide for all singleton and doubleton sets. Since $\phi$ and $\psi$ are quadratic, it follows that $\phi =\psi$.
\end{proof}

\bigskip
\noindent\textbf{\large Acknowledgement.}\enspace
The author thanks the referee for correcting mistakes in the original manuscript and making suggestions that greatly improve this paper.

\end{document}